\documentclass[a4paper,onecolumn]{quantumarticle}
\pdfoutput=1

\usepackage{graphicx}
\usepackage{textcomp}
\usepackage{xcolor}
\usepackage{listings}
\usepackage{hyperref}
\usepackage{cleveref}
\usepackage{color,soul}
\usepackage{framed}

\definecolor{keywordcolor}{RGB}{100,0,100}
\definecolor{commentcolor}{RGB}{50,100,50}
\begin{document}

\title{Introducing CQ: A C-like API for Quantum Accelerated HPC}

\author[0,1]{Mateusz Meller}
\email{Mateusz.Meller@ed.ac.uk}
\author[1]{James Richings}\orcid{0000-0003-0895-4063}
\author[0]{Vendel Szeremi}
\author[1]{Oliver Thomson Brown}\orcid{0000-0002-5193-8635}

\affil[0]{STFC Hartree Centre, Daresbury, United Kingdom}
\affil[1]{EPCC, University of Edinburgh, Edinburgh, United Kingdom}

\maketitle
\begin{abstract}
In this paper we present CQ, a specification for a C-like API for quantum accelerated high-performance computing (HPC), as well as CQ-SimBE, a reference implementation of CQ written in C99, and built on top of the statevector simulator QuEST \cite{jones19,QuEST}. CQ focuses on enabling the incremental integration of quantum computing into classical HPC codes by supporting runtime offloading from languages such as C and Fortran. It provides a way of describing and offloading quantum computations which is compatible with strictly and strongly typed compiled languages, and gives the programmer fine-grained control over classical data movement. The CQ Simulated Backend (CQ-SimBE) provides both a way to demonstrate the usage and utility of CQ, of quantum computing as an accelerator, and a space to experiment with new features such as support for analogue quantum computing, multiple QPUs, or a native Fortran interface. Both the CQ specification and CQ-SimBE are open-source, and available in public repositories online \cite{CQ-Spec,CQ-SimBE}.
\end{abstract}

\section{Introduction}

As quantum computing hardware continues to improve it becomes important to consider the future role of quantum computing as an accelerator of classical HPC, and indeed, this has been done at every level from hardware to system software, to algorithms \cite{BMH17, Mohseni24, SSRW23, A-YCRK24}. If there is a lesson to be learned from efforts to introduce classical accelerators to the HPC ecosystem however, it is that it is never too early to begin considering and experimenting with programming models for novel hardware, and one can never have too many. Arguably, efforts to exploit GPUs at scale for scientific computing are still ongoing, and we anticipate quantum accelerators being considerably harder to integrate. 

In this paper we present \emph{CQ} \cite{CQ-Spec}, a C-like API and programming model for hybrid quantum HPC. We will begin by considering the current state of the art in this area, and highlight the gap which we believe CQ fills. The paper is then split into two broad sections. In the first, we shall consider the design of CQ as a specification, discussing the design goals and decisions that shaped it, and highlighting key parts of the API. In the second, we discuss a reference implementation built on top of the \emph{QuEST} statevector simulator \cite{jones19}. The reference implementation, \emph{CQ-SimBE} \cite{CQ-SimBE} (for \emph{Simulated BackEnd}), simulates runtime offloading of quantum kernels using the CQ API and allows users to implement and demonstrate hybrid quantum applications. Crucially, this implementation supports Message-Passing Interface (MPI) communication, putting realistic constraints on host-device communication and enabling `what-if' studies for multi-device scenarios. Finally, we will discuss our roadmap for future work on CQ and CQ-SimBE.

\subsection{State of the Art}

Programming models and software frameworks for quantum computing are, like the hardware, in a nascent state. Emphasis has been placed on ease-of-use, and compatibility with near-term hardware which is often hosted in the cloud. As a result, early vendor-supplied frameworks such as Qiskit (IBM), Pennylane (Xanadu), and TKET (Quantinuum) have provided python interfaces and assumed a loose coupling between the quantum and classical parts of the code \cite{qiskit24,pennylane22,Sivarajah21}.

When it comes to integration with HPC, much of the focus has been on intelligent orchestration of heterogeneous compute resources through workflow managers such as Tierkreis \cite{sivarajah22} and Covalent \cite{cunningham23}, with a focus on integration with Slurm \cite{shehata26,esposito25}. Recently a cooperation of many vendors and research groups culminated in the development of Quantum Resource Management Interface (QRMI) \cite{badts_examining_2026}, which is integrated with major workload managers (Slurm, PBS, LSF, Grid Engine, and Kubernetes). There are however projects which focus on a lower level of integration, directly into user or library code, and providing C/C++ interfaces rather than interpreted language interfaces. Notable projects include XACC \cite{xacc_2020}, CUDA-Q \cite{cuda-q}, and the Munich Quantum Software Stack -- in particular the \emph{Quantum Programming Interface}, and the \emph{Quantum Device Management Interface} \cite{Kaya24,qdmi}. Aside from these developments, there has been a recent focus on extending existing standards such as OpenCL \cite{vazquez-perez24}, OmpSs-2 programming model \cite{dobler2025}, or using \emph{pragma} directives to manage quantum computation \cite{gazda23}. In a similar spirit, there is also an experimental implementation of quantum task offloading with the OpenMP API \cite{lee23}, though it suffered from a lack of a C-like way to describe quantum tasks -- indeed, that was part of the original inspiration for CQ. Recently, the idea of extending OpenMP to quantum computing has been revisited, culminating in OpenMP-Q \cite{swastik2025}, which generates Python scripts from the OpenMP target region and transfers the data via POSIX pipe. As noted by the authors, however, the proposed specification is language-agnostic and there is still a need to describe quantum computations natively in C/C++ and Fortran. Furthermore, this approach would ultimately need to be adopted and supported by the OpenMP Architecture Review Board, a process which can take many years. Finally, one can argue that the adoption of GPUs as OpenMP target stems from the their inherent massive parallelism. With QPUs the picture is less clear and, indeed, offloading to quantum devices might be a poor fit for OpenMP programming model.

The programming models we have discussed so far are for the digital modality of quantum computing (also known as the ``gate model''). The situation is similar in the analogue quantum computing modality, which includes quantum annealing, albeit with much less support overall. There is a dominance of python-based, loosely coupled programming models. For further insight on HPC-centric programming models for analogue quantum computing, we refer the interested reader to our recent article on the topic \cite{meller25}.

\subsection{Contributions}
The gaps that we have identified in the field, and attempt to fill with CQ are programming models with low toolchain complexity, embedded in existing HPC programming ecosystem, and which empower the programmer to precisely control the timing of offload to the quantum device and the synchronisation of results back to the classical host. Moreover, compared to compiler-oriented solutions we focus on providing a lightweight, runtime-based solution with enhanced portability and interoperability so vital for the classical HPC ecosystem. To this end, the interface presented in this work and examples included in the associated repository \cite{CQ-SimBE}, highlight that our solution advances state-of-the-art, in particular, providing constructs for quantum-accelerating existing scientific codes with minimal friction.

\section{CQ Specification}

The design goals of the CQ specification are:
\begin{itemize}
    \item A strictly and strongly typed interface, with maximal compatibility with the compiled languages commonly used in the HPC setting (C/C++/Fortran).
    \item Minimising toolchain complexity, and relying as much as possible on runtime libraries and standard compilers.
    \item Giving the user full control over when and how quantum kernels and classical data are transferred between host and device.
    \item Providing flexibility to implementers -- the specification only describes the user interface, and should be considered minimal in the sense that extending the interface is always permitted. 
\end{itemize}

The aim ultimately being to support the incremental acceleration of existing HPC codes, leveraging quantum resources. The CQ specification is hosted on a public GitHub repository and available under an MIT License \cite{CQ-Spec}. 

\subsection{Hardware Model}

\begin{figure}[htb!]
    \centering
    \includegraphics[width=0.85\linewidth]{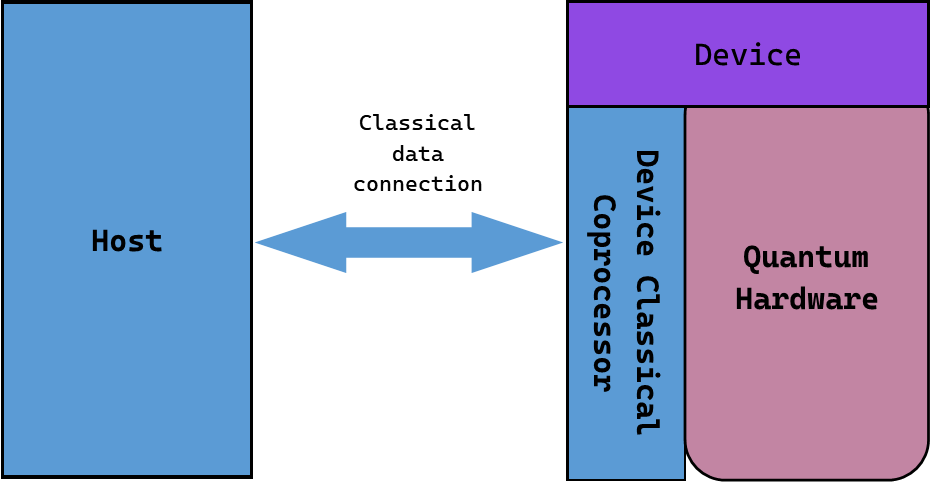}
    \caption{The hardware model assumed by CQ. A classical host shares classical information with a classical coprocessor which has direct access to quantum hardware. CQ provides an interface to synchronise data between host and device, and to define quantum circuits, but does not specify what it means to run a quantum circuit on the device. This is considered an implementation detail.}
    \label{fig:hwmodel}
    % \Description{This figure presents the hardware model assumed by CQ. On the left there is a box representing the classical `Host' process. On the right is a box representing the quantum `Device'. The Device box is further sub-divided into a box representing the classical coprocessor, and another representing the quantum hardware. There is a bidirectional arrow between the Host and Device boxes labelled `Classical data connection'.}
\end{figure}

CQ assumes the hardware model pictured in \cref{fig:hwmodel}. There is a classical host and a device which consists of a quantum computer (or quantum processing unit -- QPU for short), and a classical co-processor. The classical co-processor is assumed to have the ability to allocate and manipulate quantum resources. Importantly, it is also assumed that the classical co-processor is capable of running compiled binaries i.e. it is a standard CPU architecture of some kind, and ideally has access to standard libraries although this could be overcome with static compilation. It is \emph{not} assumed that the device classical co-processor is as powerful as the host. It is assumed that a classical data connection exists between the device and host -- it is not strictly necessary for the device and host to be physically separated, but it should be possible.

It is assumed that the quantum part of the device is able to allocate and deallocate quantum resources at the hosts behest, initialise qubits to arbitrary classical states, perform a series of transformations on the qubits forming a quantum computation, and measure the qubits to obtain a classical result. It is assumed that the host does \emph{not} have these capabilities. There is an assumption that there may be more than one quantum device available (defined as \emph{quantum backend}), however each device will be exclusively controlled by an individual quantum kernel -- only one quantum kernel may run at once on any specific device.

\subsection{Specification Rationale}
The hardware model described above mimics the established hardware model of GPU-accelerated systems, enabling us to apply lessons learned from the classical accelerators space and optimisation of those systems. Indeed, some of the bottlenecks include data transfer, optimal problem partitioning, device resource saturation, and resource utilisation. Similarly when dealing with quantum hardware, it is expected that the data transfer will negatively impact the performance. The impact can be substantial, especially in algorithms which require frequent bidirectional communication between classical and quantum compute resources, such as quantum-enhanced parallel tempering \cite{ferguson_methods_2026, marshall_quantum-enhanced_2026}, and quantum machine learning methods (Ref.~\cite{gokhale2026} provides a recent review of state of the art). The standard way to alleviate these challenges is to give user control over resource management and minimise the number of operations that would trigger transfer of data, cleanly communicating within the interface that it is an undesirable action. Given the engineering challenges around scaling of quantum devices, it is easy to saturate the device capabilities either in terms of qubits used or available program duration (usually referred to as quantum circuit depth). As such, it is important to provide user with interface that communicates device constraints, at the same time, enabling control over the device's resources and sub-routines (also referred to as kernels) that are executed on the device. Finally, to maximise resource utilisation, common optimisation techniques require problem partitioning and overlapping computation across multiple heterogenous compute resources, such that, ideally, no unit is idle. This requires an interface that allows for asynchronous communication and offloading. It is those issues and classical solutions that guided the shape of CQ specification. 

\subsection{Resource Management}

CQ includes functions to allocate and deallocate quantum resources on the quantum device presented in \cref{lst:cq-res-mgt}:

\begin{lstlisting}[label={lst:cq-res-mgt}, language=C, caption={CQ-defined resource management functions. Users can allocate and free either a single qubit or a collection of qubits which can be modelled as a quantum register.}, captionpos=b]
int alloc_qubit(qubit ** qubit_handle);
int alloc_qureg(qubit ** qureg_handle, size_t nqubits);
int free_qubit(qubit ** qubit_handle);
int free_qureg(qubit ** qureg_handle);
\end{lstlisting}

The simplest form of resource management, allocates a single qubit (\lstinline$alloc_qubit$). A more general way to operate on a whole set of quantum qubits (or a quantum register, `qureg' for short) can be achieved with \lstinline$alloc_qureg$.
These functions are allowed on the host for two reasons -- first, there is an assumption that allocating or deallocating quantum resources may be a lengthy process, and therefore it is desirable for the programmer to do this once at the start and end of their program, and reuse allocated resources in between. Secondly, these functions are also responsible for creating or destroying the classical handle associated with quantum resource.

The opaque datatype \lstinline$qubit$ may contain any information required by the implementer to keep track of quantum resources on the device, and which may then be supplied to a quantum kernel. The \lstinline$qubit$ datatype should also obey C pointer semantics, such that the $i^{\mathrm{th}}$ qubit from a register \lstinline$qr$ may be accessed using \lstinline$qr[i]$ or even \lstinline$*(qr+i)$ from within a quantum kernel. Finally, note that the allocated qubits are currently assumed to be `logical' qubits. CQ does not currently have any conception of noise or error correction.

\subsection{Quantum Kernels}

\begin{lstlisting}[label={lst:cq-state-setters}, language=C, caption={Device-only set of quantum state setters. User can set either an individual qubit, a quantum register to an integer state (interpreted as a bitstring), or a quantum register based on the state of a classical register.}, captionpos=b]
int set_qubit(
    qubit qubit_handle, 
    cstate cs);
int set_qureg(
    qubit * qureg_handle, 
    const unsigned long long STATE, 
    const size_t NQUBITS);
int set_qureg_cstate(
    qubit * qureg_handle,
    cstate const * const
      CSTATE_REGISTER,
    const size_t NQUBITS);
\end{lstlisting}

CQ specifies the opaque datatypes \lstinline$qkern$ and \lstinline$pqkern$ for quantum kernels and parameterised quantum kernels respectively, however the precise definition of these is implementation defined. In CQ-SimBE, they are function pointers. Separating quantum kernels and parameterised quantum kernels in this way allows implementers to handle them differently if necessary -- again in CQ-SimBE this means for example that they have different function signatures. Within a quantum kernel, the user may use the device-only functions including quantum state setters, quantum measurements, and quantum gates. The quantum state setters use the interface outlined in \cref{lst:cq-state-setters}.
These functions are limited to setting qubits to \emph{classical} states, mirroring a limitation of real quantum hardware. In general preparing a specific quantum state requires the design and construction of a state preparation circuit. The opaque \lstinline$cstate$ type used here may be any datatype capable of holding a classical binary state, such as a \lstinline$bool$ or a \lstinline$short int$.

The general pattern expected of a quantum kernel is state initialisation, state transformation, then measurement. CQ has separate interfaces for measurements which will and will not be synchronised back to the host. The synchronised measurement interface is presented in \cref{lst:cq-measure}.\\

\begin{lstlisting}[label={lst:cq-measure}, language=C, caption={CQ interface for synchronised measurements. User can measure either a single qubit, a whole quantum register or they can select which qubits in the register are to be measured.}, captionpos=b]
int measure_qubit(qubit * qb, cstate * result);
int measure_qureg(qubit * qr, const size_t NQUBITS, cstate * results);
int measure(
    qubit * qr, const size_t NQUBITS, size_t const * const TARGETS,
    const size_t NTARGETS, cstate * results);
\end{lstlisting}
The unsynchronised device measurement interface uses \lstinline$dmeasure$ variants. The difference between the \lstinline$measure$ and \lstinline$dmeasure$ interface is only the expectation that the result buffer is local to the device coprocessor in the device measurement variant, however separating these interfaces allows the programmer to better express intent, and gives CQ implementers the opportunity to include synchronisation operation only in the host measurement functions. The \lstinline$measure$ function provides the most general interface, while the other two are specialisations for common uses.

\subsubsection{Gate Sets}

As noted in the introduction, the focus of CQ is on managing the offload of quantum kernels and synchronisation of classical data between host and device, rather than on the design of quantum kernels themselves. With that in mind, we have taken a modular approach to specifying the quantum operations available within a kernel. For gate-model quantum computing any universal gate set should do, but particular choices may be convenient for particular applications or hardware. Currently, the only gate set defined in the specification is the OpenQASM standard gate set \cite{qasm-gates}. We have included the interface for a one- and two-qubit gate from this set below for illustrative purposes (\cref{lst:cq-gates}). \\

\begin{lstlisting}[label={lst:cq-gates}, language=C, caption={Example interface for a single and two-qubit gate. Qubit operands are passed as a handle, followed by gate specific parameters such as a rotation angle.}, captionpos=b]
int hadamard(qubit * target);
int cphase(qubit * ctrl, qubit * target, const double THETA);
\end{lstlisting}

\subsection{Kernel Offloading and Synchronisation}

Finally, and perhaps most importantly, CQ includes an interface for offloading and synchronising quantum kernels in a variety of ways. These are all \emph{host} operations. The name of each executor in CQ consists of a letter prefix, and the suffix \lstinline$_qrun$. The prefix letters are presented in \cref{tab:exec_prefix}.

\begin{table}[htb!]
    \centering
    \begin{tabular}{c|p{0.85\textwidth}}
        Letter & Meaning \\
        \hline
        s & \emph{Synchronous}. This executor will block until the quantum kernel completes, and the result buffer is synchronised back to the host. \\\\
        a & \emph{Asynchronous}. This executor will return immediately. The result buffer is not guaranteed to be synchronised back to the host until an appropriate synchronisation function is called. \\\\
        m & \emph{Multi-shot}. This executor will run the quantum kernel multiple times. The result buffer should be large enough to hold measurement results from all shots. \\\\
        b & \emph{Backend}. This executor will offload the quantum kernel to the specified backend. Here by backend we define quantum device which is identified by \emph{BackendId}. Executors specifying different backends allow running multi-QPU workflows in a similar spirit to multi-GPU workflows.  \\\\
        p & \emph{Parameterised}. The quantum kernel is parameterised, so this executor accepts a \lstinline$pqkern$ and a parameter pack.
    \end{tabular}
    \caption{Prefix codes for the \lstinline$qrun$ executor functions.}
    \label{tab:exec_prefix}
\end{table}
All relevant prefix combinations are included in the specification for a total of 16 different executors. The synchronicity prefix \emph{s} or \emph{a} is included in every executor, while \emph{m}, \emph{b}, and \emph{p} are only included when relevant. That is to say, the omission of \emph{m} indicates the executor is single-shot, the omission of \emph{b} indicates it will be executed on the default backend, and the omission of \emph{p} indicates it accepts an unparameterised quantum kernel.

The simplest executor is the unparameterised, single-shot, synchronous, default backend executor, depicted in \cref{lst:cq-sqrun}.\\

\begin{lstlisting}[label={lst:cq-sqrun}, language=C, caption={CQ declaration of a synchronous executor, which accepts a quantum kernel, a quantum register and performs single-shot, unparameterised execution.}, captionpos=b]
int s_qrun(
    qkern kernel, qubit * qr, const size_t NQUBITS, 
    cstate * const result, const size_t NMEASURE);
\end{lstlisting}
Its parameters are common to all the executors. An important note is the relationship between the result buffer, the parameter \lstinline$NMEASURE$, and the quantum kernel. The \lstinline$cstate$ result buffer \emph{must} be large enough to hold at least \lstinline$NMEASURE$ measurement results, and \lstinline$NMEASURE$ \emph{must} be greater than or equal to the maximum number of measurement results which will be stored in a single execution of the quantum kernel. All this really means in this case is that the result buffer should be large enough to hold all the measurements, and \lstinline$NMEASURE$ should correspond to the allocated length of the result buffer, but there are a number of complicating factors. First, the kernel may of course perform a number of device-only measurements which will not be stored in the main result buffer -- these do \emph{not} count towards \lstinline$NMEASURE$ (as it is used solely to inform host-device synchronisation). Second, the kernel may make measurements conditionally, and so it is not guaranteed that the entire result buffer is actually used on successful completion of a quantum kernel, but of course space must still be allocated and \lstinline$NMEASURE$ should reflect that. Finally, in the multi-shot executors the \lstinline$const size_t$ parameter \lstinline$NSHOTS$ is introduced. In that scenario \lstinline$NMEASURE$ should still correspond to the maximum number of synchronised measurements from a \emph{single} execution, and the expected allocated size of the result buffer is \lstinline$NMEASURE$ \(\times\) \lstinline$NSHOTS$ \(\times\) \lstinline$sizeof(cstate)$ bytes.\\

\begin{lstlisting}[label={lst:cq-aqrun}, language=C, caption={CQ declaration of an asynchronous executor, which accepts a quantum kernel, a quantum register and performs single-shot, unparametrised execution.}, captionpos=b]
int a_qrun(
    qkern kernel, qubit * qr, const size_t NQUBITS, 
    cstate * const result, const size_t NMEASURE, 
    struct exec * const exec_handle);
\end{lstlisting}
The asynchronous executors such as the one presented in \cref{lst:cq-aqrun}, include an additional execution handle parameter which allows the programmer to later synchronise the offloaded kernel. The synchronisation interface is depicted below (\cref{lst:cq-sync}),
\\
\begin{lstlisting}[label={lst:cq-sync}, language=C, caption={Set of CQ-defined synchronisation functions. User can either synchronise the state of the executor mid-run, wait until quantum kernel completes execution, or halt the execution of the quantum kernel.}, captionpos=b]
int sync_qrun(struct exec * const exh);
int wait_qrun(struct exec * const exh);
int halt_qrun(struct exec * const exh);
\end{lstlisting}
where in each case \lstinline$struct exec * const exh$ is a handle corresponding to a specific execution -- \lstinline$struct exec$ is another opaque datatype intended to contain whatever the implementer needs to facilitate synchronisation. It should be safe to reuse a \lstinline$struct exec$ once the execution has completed. The intended usage of opaque type \lstinline{struct exec} is via pointers. As such, CQ defines standard allocation and deallocation functions and minimal set of data access functions to interact with execution handles (see \cref{lst:cq-exec} in appendix \ref{appendix:exec_interface} for details).  The first synchronisation function, \lstinline$sync_qrun$, immediately synchronises the results buffer to the host, but does not guarantee the completion of the execution. The second, \lstinline$wait_qrun$ blocks until the execution is complete and the final results buffer is synchronised to the host. Finally, \lstinline$halt_qrun$ does the same as \lstinline$wait_qrun$ but additionally \emph{requests} early completion of the execution. This may, for example, be used in conjunction with \lstinline$sync_qrun$ to end a multi-shot execution early if the result is identified from partial results, however the precise details and the granularity at which \lstinline$halt_qrun$ operates is left to the implementation.

Operations defined above enable a set of potentially useful optimisations in the context of hybrid algorithms. The simplest and perhaps the most common execution pattern allows for overlapping computation across host and the device. Using asynchronous executors, host can offload quantum workload and continue classical computation, until it requires results from the device. Then it calls \lstinline$wait_qrun$ to synchronise and retrieve data from completed kernel. Another pattern, enabled by \lstinline$halt_qrun$, could be used to implement an early-exit, where both host and the device do work, but depending on the condition from classical processing, quantum job is cancelled and results gathered up to this point are synchronised back to the host. We note, that this pattern is expected to benefit systems of the future, where computation on the fault-tolerant quantum devices can take a long time to complete. Finally, if the data coming from the quantum device requires heavy classical post-processing, \lstinline$sync_qrun$ allows for some kind of pre-fetching, improving resource utilisation and reducing time-to-solution. These are only some of the example uses of the introduced routines. We expect that users will find benefits coming from those constructs as the field of hybrid quantum-classical workflows matures.

\section{CQ-SimBE Reference Implementation}

In order to assess the practicality of the interface we have created a reference implementation in C99, built on top of the statevector simulator QuEST \cite{jones19}, which takes the place of the quantum computer. The reference implementation is called the \emph{CQ Simulated Backend} or \emph{CQ-SimBE}, and it is available in a public GitHub repository under an MIT license \cite{CQ-SimBE}. As well as QuEST, CQ-SimBE currently depends on POSIX threads, Unity test framework \cite{unity} and, optionally, MPI. 

\begin{figure}[htb!]
    \centering
    {\setlength{\fboxsep}{0.075\textwidth}
    \fbox{\includegraphics[width=0.75\textwidth]{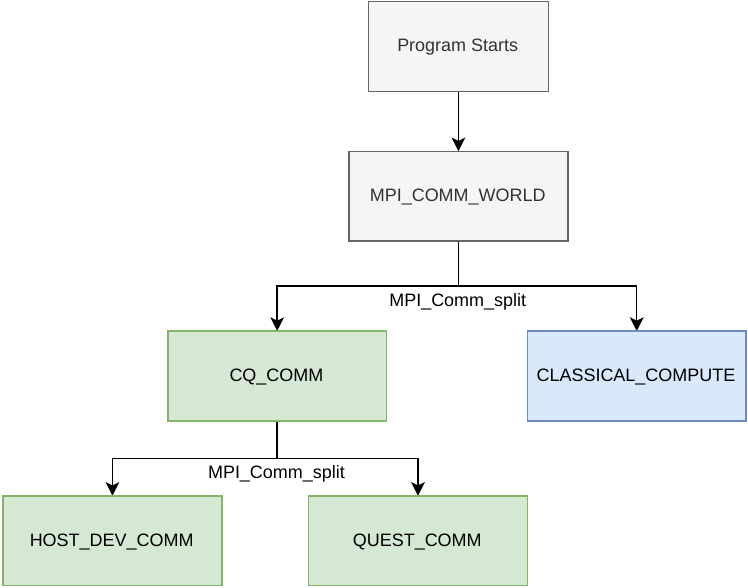}}}
    \caption{The MPI communicator splitting as implemented by CQ-SimBE. CQ-SimBE supports applications where the user owns MPI communicator. In that scenario, user should split the communicator into one responsible for offloading and one used for classical computation. Then internally, CQ splits the received communicator further to handle host-device communication and distributed statevector simulation (using QuEST). To simplify writing hybrid programs, we provide additional C preprocessor macros \lstinline{CQ_PROG_BEGIN} and \lstinline{CQ_PROG_END}, to ensure this block of code is executed by CQ only.}
    \label{fig:mpi_comm}
    % \Description{This figure presents the MPI communicators structure supported and assumed by CQ. On the top there is a box showing program entry. Below, we have box representing global MPI communicator. From this box, two arrows go towards other boxes labeled `CQ\_COMM' (on the left), representing CQ MPI sub-communicator and `CLASSICAL\_COMPUTE' (on the right) representing MPI communicator used for classical computation. Below `CQ\_COMM' box, there are additional two boxes, showing further splitting of the communicator to Host-device communicator responsible for this communication and on the right, `QUEST\_COMM', responsbile for distributed statevector simulation.}
\end{figure}

\begin{figure*}[htb!]
    \centering
    {\setlength{\fboxsep}{0.075\textwidth}
    \fbox{\includegraphics[width=0.75\textwidth]{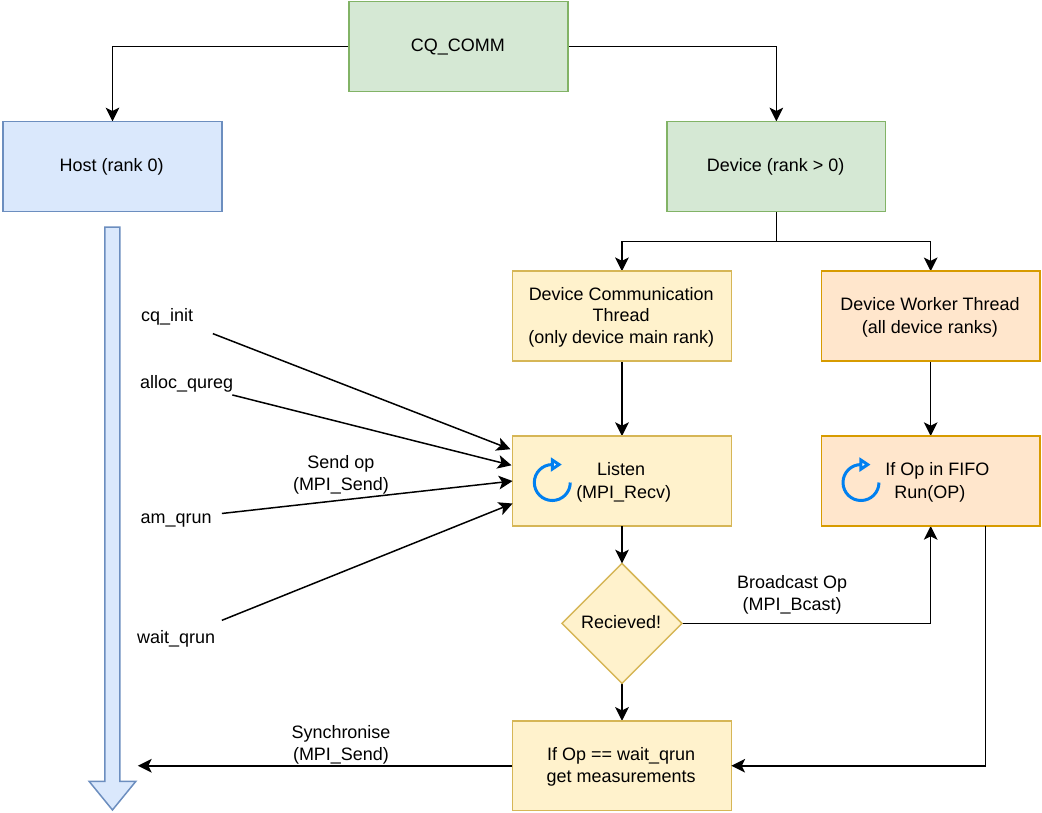}}}
    \caption{Implementation details of the host-device communication and quantum operations execution. Within CQ sub-communicator, the main rank (e.g. rank 0) is considered the host, which executes host only operations and other classical code (where blue arrow represents execution flow). The rest of MPI processes (e.g. ranks $>$ 0) are considered to belong to the device. The main device rank (e.g. rank 1) spawns two POSIX threads. The first one is responsible for communication with the host and awaits for incoming messages. The other thread is a worker thread (spawned by all device ranks) which executes device specific operations (i.e. runs distributed statevector simulation). If the communication thread receives an operation from the host (along with relevant data), the operation (and data) are broadcasted to all the workers. For a set of synchronisation operations (e.g. \lstinline{wait_qrun}), the results of the simulation are fetched and sent back to the host.}
    \label{fig:cq_simbe}
    % \Description{This figure presents the Implementation details of CQ-SimBE. The top box represent what happens inside CQ sub-communicator. The left box represents the `host' and below the box there is an arrow, showcasing execution flow of example host-only operations. On the right, there is a box representing the `device'. From the `device', two arrows go to two boxes below which create two columns, representing separate POSIX threads. The left column, starts with the box showing that it is a `device communication thread'. The box below it shows that it executes a loop, listening for messages from the host. Below is a diamond box which represents and event, that the message was received. The box at the bottom, represents a condition, showing what happens if the `host' requests synchronisation. The rightmost column, depicts `device worker thread'. Below the first box, there is another box which represents that the worker thread operates in the loop, executing quantum operations in the queue.}
\end{figure*}

CQ-SimBE can be built in two different modes of operation, specifying how the offloading to a quantum computer is simulated. The first mode implements host-device communication using two POSIX threads, one to act as the host and one as the device.
% currently simulates offloading to a quantum computer using two POSIX threads, one to act as the host, and one as the device.
This restriction to shared memory naturally simplifies the implementation of the executors and data synchronisation functions, however, CQ-SimBE has been designed with message-passing in mind. The second mode implements host-device communication with a mix of MPI and POSIX threads as depicted in \cref{fig:mpi_comm} and \cref{fig:cq_simbe}. In this scenario the main MPI communicator is split into two sub-communicator; one responsible for host-device communication and data transfer, and one responsible for on-device execution and statevector simulation (\cref{fig:mpi_comm}). On the device, the main process is responsible for receiving data and control operations from the host and broadcasting those to the rest of device-assigned worker processes. To implement this logical separation of concerns, the device process spawns two POSIX threads (\cref{fig:cq_simbe}). The first thread listens and awaits for messages from the host and then distributes them to other workers. The second processes operations across all workers and in practice executes quantum kernels. Then the results are collected across the workers and synchronised back to the host via communication thread. This design was motivated by a desire to introduce cross-device communication in future work.
% and we expect to extend the implementation to optionally use POSIX threads or MPI to separate the host and device.

To facilitate flexible adoption within existing MPI-based applications where the user owns the MPI communicator, we provide an additional routine: \lstinline{cq_init_custom_mpi_comm}, which is intended to setup internal state of the CQ and handle subsequent creation of valid sub-communicators required for CQ operation.

In both modes of operation, we make the use of the quantum kernel registration interface to construct a map of (local) function pointers, allowing us to simulate remote procedure calls \cite{nelson1981, birrell1984}. The `device' thread maintains a linear FIFO queue of control operations, which can be added to by the host. Function pointers for the control operations are stored in an array, so the host need only transfer a simple \lstinline$int$ to the device to request a control operation. More complex operations, such as quantum kernels, do of course require additional data to be transferred such as the key corresponding to a registered quantum kernel, or the number of qubits to be allocated. 

The CQ \lstinline$cstate$ datatype is compile-time defined as a \lstinline$short int$ in CQ-SimBE, allowing negative values to be used to flag invalid values. The \lstinline$qubit$ datatype is implemented as a struct (\cref{lst:cq-qubit}),
where \lstinline$registry_index$ selects an allocated QuEST statevector from a register held on the device, and \lstinline$offset$ indexes into that register. The final data member \lstinline$N$ holds the number of qubits in the quantum register, and is not strictly necessary, but is helpful for debugging and testing.

\begin{lstlisting}[label={lst:cq-qubit}, language=C, caption={Example definition of a qubit type used by the CQ-SimBE. Each qubit tracks the parent registry, its offset and the number of qubits in the register.}, captionpos=b]
typedef struct qubit {
    size_t registry_index;
    size_t offset;
    size_t N;
} qubit;
\end{lstlisting}

\begin{lstlisting}[label={lst:cq-qkerns}, language=C, caption={Declaration of CQ-compliant unparametrised and parametrised kernels respectively. The parametrised kernel differs from unparametrised version by accepting extra void pointer to arbitrary structure and different kernel tracking type.}, captionpos=b]
cq_status user_qkern(
    const size_t NQUBITS, qubit * qreg, 
    cstate * creg, struct qkern_map * reg);
    
cq_status user_pqkern(
    const size_t NQUBITS, qubit * qreg, 
    cstate * creg, void * params, struct pqkern_map * reg);
\end{lstlisting}

Quantum kernels in CQ-SimBE are defined in \cref{lst:cq-qkerns},
where \lstinline$NQUBITS$ is the total number of qubits allocated in \lstinline$qreg$, \lstinline$creg$ is the buffer for results which will be synchronised back to the host, and in the parameterised quantum kernel, \lstinline$params$ is the user-defined parameter pack. The \lstinline$registration$ kernel is part of the kernel registration process -- when it is not \lstinline$NULL$ the struct is populated with a key for the device kernel register by a macro the user is required to insert at the top of their kernels. While this approach adds clutter to user quantum kernels, it also allows us to return early after registering the kernel without running it. Importantly, only the entry point for the user kernel needs to be registered since we already rely on all functions existing locally in the single program, multiple data (SPMD) model of MPI.

\subsection{Fortran Interface}

In line with the design goals of CQ, we have also implemented a Fortran 2008 interface in CQ-SimBE, enabling quantum kernel definition and offload from Fortran programs. By defining an abstract interface to a function, the Fortran API enables users to define the quantum kernels natively, using supplied device-specific functions. In addition to this, the interface provides module function definitions for each CQ function, allowing to manage the CQ environment, request quantum resources from the host, register quantum kernels and offload kernels to the device. Below, in \cref{lst:cq-fortran}, we provide few examples of the CQ functions mapped to Fortran interface,
\\
\begin{lstlisting}[label={lst:cq-fortran}, language=Fortran, caption={Excerpt of CQ-compliant Fortran interface which includes functions for environment and resource management as well as quantum kernel registration, and kernel offloading.}, captionpos=b]
cq_init(VERBOSITY)
cq_finalise(VERBOSITY)
alloc_qureg(qrp, N)
free_qureg(qrp)
sync_qrun(ehp)
create_exec_handle(ehp)
free_exec_handle(ehp)
register_qkern(kernel)
sm_qrun(kernel, qrp, NQUBITS, crp, NMEASURE)
am_qrun(kernel, qrp, NQUBITS, crp, NMEASURE,ehp)
\end{lstlisting}
where \lstinline$VERBOSITY$, \lstinline$NQUBITS$, \lstinline$NMEASURE$, \lstinline$NSHOTS$, are \lstinline$integer$ (of kind 8), \lstinline$crp$ is an \lstinline$integer$ array, \lstinline$kernel$ is of type \lstinline$procedure(qkern)$, \lstinline$qrp$ and \lstinline$qhp$ are derived types of \lstinline$qubit$, and \lstinline$cq_exec$ respectively, which are \lstinline$c_ptr$ wrappers. Given one-to-one mapping with C interface, for complete outline of the Fortran interface we refer user to Ref.~\cite{CQ-Spec} for brevity.

One additional difference is the kernel registration macro, where the user now must pass the kernel name explicitly, as the C preprocessor macro used to extract this information in the C interface is not available in Fortran. An example of a quantum kernel written in Fortran is included in appendix \ref{appendix:fort_interface}.

\subsection{Analogue Quantum Computing Interface}
To allow targeting applications such as Hamiltonian simulation or quantum annealing, we have extended CQ-SimBE with an experimental analogue quantum computing interface -- \emph{cq-analog}. The module focus is on defining pulses and selecting channels which can execute those pulses to manipulate qubits in the quantum registers. We present the details of the analogue extension below.\\

\begin{lstlisting}[label={lst:cq-analog-env}, language=C, caption={Set of CQ analogue environment-managing functions. To use analogue operations user has to enable analogue mode on the host side, first. Then inside the quantum kernel definition, user can toggle quantum register as analogue compatible.}, captionpos=b]
int enable_analog_mode(int mode);
int enable_analog_qreg(qubit *qr);
int disable_analog_qreg(qubit *qr);
\end{lstlisting}

\subsubsection{Host Operations}
The module adds one extra host operation -- \lstinline$enable_analog_mode(int mode)$ (\cref{lst:cq-analog-env}), which needs to be called before registering the quantum kernel. The function initialises the device parameters and ensures that the analogue operations can be executed correctly. The function takes one argument -- \lstinline$mode$, which determines in which analogue mode the device operates. In the provided implementation, the available modes are \lstinline$ISING$ and \lstinline$XY$, referring to the different Hamiltonians that the device implements. In general, \lstinline$mode$ is device-specific and should be provided by the vendor.

\subsubsection{Device Operations}

Within quantum kernels, there is a set of new device operations. First of all, the user has to enable the quantum register to operate in analogue mode (\cref{lst:cq-analog-env}). The function takes as an argument a quantum register handle. This is crucial for the device to register and correctly direct the analogue operations. The next operation, \lstinline$disable_analog_qreg$ (\cref{lst:cq-analog-env}) is a clean-up operation, which takes the handle to the quantum register as an argument and frees the resources, unlocking the register for other tasks. Currently, CQ allows mixing of the different kinds of device operations. That is, if the quantum register is enabled in analogue mode, the user can both execute digital (gates) operations on the qubits and analogue pulses. The rationale for this is that for some applications, hybrid digital-analogue processing is beneficial \cite{parra-rodriguez20}. However, if this is undesirable from the hardware vendor's perspective, it is up to the implementer to prohibit it.
\\
\begin{lstlisting}[label={lst:cq-analog-channel}, language=C, caption={Analogue-specific CQ channel management functions. Users can request channel which, depending on channel type can either operate on set of qubits or single target qubit. If the channel supports local targetting (as opposed to global), user further can retarget the channel.}, captionpos=b]
int get_channel(channel *ch, int type, qubit *qr, qubit *target);
int retarget_channel(channel *ch, qubit *new_target);
\end{lstlisting}

Having enabled the device in analogue mode, the user can request a handle to a channel which operates on qubits (\cref{lst:cq-analog-channel}). The channel can have a \lstinline$GLOBAL$ or \lstinline$LOCAL$ addressing mode, i.e. either it operates on all qubits within the register or targets specific qubits. All the limitations on both the number of available channels and targeting specifics are up to the implementer. The module does not make any assumptions about it. If the channel is addressing qubits in the \lstinline$LOCAL$ mode, users can re-target the channel to operate on a selected qubit.

The next set of device-specific operations are the pulses. These are presented in \cref{lst:cq-analog-pulse}. The \lstinline{pulse} is another opaque datatype which should hold any information required to define pulses with corresponding waveforms. In the presented implementation, the \lstinline$pulse$ datatype is a struct which contains arrays of frequencies, phases and optionally detuning, which is useful on the neutral atom platforms. Those three arrays specify the shape of the pulse throughout its duration.  Moreover, the \lstinline$pulse$ struct has a duration and number of samples as members for the ease of further processing. The units of time are up to the device; however, in our example implementation, duration is assumed to be in nanoseconds.  After a pulse has been initialised, its parameters should be set using one of the supplied waveforms, which are shortly described in \cref{sec:analog-waveforms}.\\

\begin{lstlisting}[label={lst:cq-analog-pulse}, language=C, caption={CQ interface for managing and execution operation on pulses. Operations also require a valid channel which drives the pulse. Additionally, pulse object should be initialised with supplied waveform functions before execution.}, captionpos=b]
int init_pulse(pulse *pulse, double duration);
int free_pulse(pulse *pulse);
int play(channel *ch, pulse *pulse);
int capture(channel *ch, pulse *pulse, int *result, int shots);
int delay(channel *ch, double dt);
int barrier(channel **ch, int num_channels);
\end{lstlisting}

With the pulse correctly prepared, the user has access to the set of functions which take channel and pulse handles and operate on qubits in the register. The play operation plays the pulse via the specified channel  applying the quantum operation defined by the pulse to all qubits targeted by the channel. Users can also capture the data from the quantum register via a pulse.  This function is provided for a more fine-grained application of measurement, in situations where the users want to have more control over the device, as the base CQ specification provides standard qubit measurement operations in the computational basis. The capture operation, aside from taking channel and pulse handles as arguments, also takes an array of results and the number of shots. It is up to the user to interpret this result array, as depending on the context, it can have a different meaning, for example, a binary number, or histogram data.
Users can delay the channel, forcing it to wait for a specified time duration. They can also synchronise channels via barrier, as applying different operations by different channels will introduce a time difference in the channel clocks. The \lstinline$barrier$ function takes as arguments the array of channel handles that are to be synchronised, and the number of channels.

The analogue module offers a function -- \lstinline$set_qubit_pos$, which accepts an array of doubles which represent the new three-dimensional coordinates for the qubits. Depending on the underlying platform, this function can serve to set the quantum register topology or for example, on a neutral atoms platform, shuffle the atoms. Generally, this functionality is heavily backend dependent, and it is up to the implementer to either prohibit or limit the frequency of calls to this function.

\subsubsection{Waveforms Library}
\label{sec:analog-waveforms}
As part of the module, we provide an example collection of the waveforms used for building the pulses. Aside from some classical waveforms, such as Gaussian, sine, cosine or saw, there is also the Blackman waveform, and we give users means of creating interpolated (via cubic splines), custom or composite waveforms. The current waveforms' arguments follow a pattern where the first argument is an array of samples, then duration, followed by wave-specific parameters such as amplitude or sigma. Adding additional vendor-specific waveforms is possible, and we recommend that users adhere to the provided format.

\section{Example Application}

As an illustrative example we include here an abridged version of a simple quantum Fourier transform (QFT) circuit implemented in CQ-SimBE, and synchronously offloaded (\cref{lst:cq-impl-qft}).\\

\begin{lstlisting}[label={lst:cq-impl-qft}, language=C, caption={Example implementation of a quantum Fourier transform (QFT) using CQ interface. The listing outlines a general structure of the CQ-written programs. First the user defines a sub-kernel which implements the QFT circuit. Then, inside the main kernel the user first has to register the kernel using a macro, set the quantum register to a valid state, call sub-kernels (if any) and finally measure the quantum register to synchronise results back to the host. On the host side, within the main function, the user has to initialise CQ and allocate resources, followed by registering the kernel on the host side. Only then can the user offload the kernel using one of the available executor commands. Finally, once the kernel finishes and results are synchronised back to the host, the results can be reported and resources can be freed.}, captionpos=b]
#include "cq.h"

void full_qft_circuit(const size_t NQUBITS, qubit * qr) {
  // Run QFT
  for (size_t i = 0; i < NQUBITS; ++i) {
    hadamard(&qr[i]); 
    for (size_t j = i+1; j < NQUBITS; ++j) {
      double angle = M_PI / pow(2, j);
      cphase(&qr[j], &qr[i], angle);
    }
  }

  for (size_t i = 0; i < NQUBITS / 2; ++i) {
    size_t j = NQUBITS - (i+1);
    swap(&qr[i], &qr[j]);
  } 

  return;
}

cq_status zero_init_full_qft(
    const size_t NQUBITS, qubit * qr,
    cstate * cr, qkern_map * reg) {
  CQ_REGISTER_KERNEL(reg);

  // Prepare state
  set_qureg(qr, 0, NQUBITS);

  // Run QFT
  full_qft_circuit(NQUBITS, qr);

  // Measure
  measure_qureg(qr, NQUBITS, cr);

  return CQ_SUCCESS;
}

int main (void) {
  const size_t NQUBITS = 10;
  const size_t NSHOTS = 10;
  const size_t NMEASURE = NQUBITS;

  cq_init(0);

  qubit * qr = NULL;
  alloc_qureg(&qr, NQUBITS);

  cstate * cr;
  cr = malloc(NMEASURE * NSHOTS * sizeof(cstate));
  init_creg(NMEASURE * NSHOTS, -1, cr);

  register_qkern(zero_init_full_qft);

  printf("Running QFT circuit "); 
  printf("on quantum device.\n");
  sm_qrun(zero_init_full_qft, qr, NQUBITS, cr, NMEASURE, NSHOTS);
  report_results(cr, NMEASURE, NSHOTS);

  free_qureg(&qr);
  free(cr);

  cq_finalise(0);

  return 0;
}
\end{lstlisting}

\newpage
When run this produces an output similar to:
\begin{lstlisting}
Running QFT circuit on QPU.
Reporting measurement outcomes:
Shot [0]: 1 0 0 1 1 0 0 0 0 0
Shot [1]: 1 1 0 0 1 0 0 1 0 1
Shot [2]: 0 0 0 0 0 0 0 0 0 1
Shot [3]: 1 1 0 0 0 0 0 1 0 0
Shot [4]: 1 0 1 1 0 0 0 1 0 1
Shot [5]: 0 0 0 0 0 0 0 1 1 1
Shot [6]: 0 0 0 0 1 1 1 0 0 1
Shot [7]: 1 1 0 0 1 0 0 1 1 1
Shot [8]: 1 1 0 1 0 0 1 0 1 1
Shot [9]: 0 0 0 1 0 1 1 0 0 1
\end{lstlisting}

In this example we have separated the actual description of the QFT circuit into the function \lstinline$full_qft_circuit$ which enables reuse of this common circuit across kernels. As that function is not an entry point for the device it does not need to be registered as a quantum kernel. The quantum kernel is \lstinline$zero_init_full_qft$ -- its signature matches the \lstinline$qkern$ definition in CQ-SimBE, and it includes the \lstinline$CQ_REGISTER_KERNEL$ macro described earlier. The actual circuit description is left to \lstinline$full_qft_circuit$, but \lstinline$zero_init_full_qft$ has responsibility for initialising the quantum register, whose state is otherwise undefined, and capturing the measurement results after running the QFT circuit. Finally, \lstinline$main$ is responsible for allocating quantum and classical resources, registering the quantum kernel, offloading it to the device, and printing results, before finally cleaning up allocated resources.

The full extended version of this example, which demonstrates both synchronous and asynchronous offload, is available in the CQ-SimBE GitHub repository \cite{CQ-SimBE}, along with an implementation of the Variational Quantum Eigensolver algorithm, and a MAXCUT solver which uses the analogue quantum computing interface. 

\section{Further Work}

First and foremost, we plan to implement CQ compliant interfaces to real quantum hardware.

\subsection{Specification}

We expect that the CQ Specification will evolve, particularly in response to our experience implementing and using it. The roadmap for the specification already includes:
\begin{itemize}
    \item After obtaining further feedback from both analogue quantum hardware experts and users, we plan to formalise the \emph{cq-analog} module from CQ-SimBE and include it in the specification.
    \item Add a \emph{cq-emulation} module of interfaces to be used when the backend is explicitly a classical emulator, not real quantum hardware.
\end{itemize}
More broadly, in future iterations of the specification we would like to add support for circuit optimisation, distinguishing between physical and logical qubits to support development of Quantum Error Correction, and inter-QPU communication. 

One area that requires further considerations is the intermixing of classical operations inside the quantum kernels. Solutions such as CUDA-Q or XACC handle this by developing a compilation toolchain, which interprets and generates appropriate instructions to run on the target hardware. Since CQ is runtime-based and procedural, it is less clear how kernels with classical instructions should be processed. The ideal long term scenario is that the standard compilers will have support for device-specific loops and branching. In a short term, however, there are three solutions to this problem. First, if we write kernels in a templated language (e.g. C++), then we can resort to constructs such as the one depicted below, in \cref{lst:cq-classic-ops1}.

\newpage
\begin{lstlisting}[label={lst:cq-classic-ops1}, language=C, caption={Templated solution to hybrid quantum-classical kernels. Note that this would require either C++ based implementation or heavy use of preprocessor macros.}, captionpos=b]
template <unsigned I>
struct loop {
  template <typename F>
  loop(F f) {
    f();
    loop<I-1>(f);
  }
};

template <>
struct loop<0> {
  template <typename F>
  loop(F) {}
};

void sub_qkern() {
  // quantum ops
}
void qkernel() {
  loop<4>(sub_qkern);
  // quantum ops
} 
\end{lstlisting}

An alternative solution and one which we encourage, is to implement vendor-specific CQ extensions. For example, one could provide additional routines similar to those detailed in \cref{lst:cq-classic-ops2}.\\

\begin{lstlisting}[label={lst:cq-classic-ops2}, language=C, caption={Custom extensions to CQ which enable user to mix classical and quantum operations within kernels.}, captionpos=b]
int cq_loop(int start, int end, int incr, fptr body);
int cq_if(bool cond, fptr body);
int cq_if_else(bool cond, fptr if_case, fptr else_case);

void loop_body(...) {
  hadamard(qr[0]);
  x(qr[1]);
  cx(qr[0], qr[1]);
}

void qkern(...) {
  set_qureg(qr, 0);
  cq_loop(0, 10, 1, loop_body);
  measure_qureg(qr, 5, cr);
}

\end{lstlisting}

Finally, the majority of frameworks for developing quantum programs which target actual hardware, construct quantum circuits by creating and modifying an object to which quantum gates are appended. In this scenario, CQ-specified procedural quantum gates can be implemented by operating on some vendor-specific circuit data structure, which is then offloaded to the actual device.

The solutions proposed above are not exclusive. Indeed, they should be employed together to create a robust CQ implementation. At this stage we leave to implementers the decision of what classical instructions should be included, as these capabilities are heavily hardware-specific and can have no support at all.

\newpage
\subsection{CQ-SimBE}

CQ-SimBE is still very much under active development and the following work is already planned:
\begin{itemize}
    \item We plan to ensure that the specification is fully implemented in CQ-SimBE -- this means adding support for offloading to multiple devices.
    \item In parallel, above point requires extension to existing MPI-based implementation to enable cross-device communication, allowing multi-device `what-if' studies.
\end{itemize}

Going forward we expect CQ-SimBE to be used as a testbed for improvements and updates to the specification.

\subsection{CQ on Quantum Hardware}
Naturally, the ultimate testing ground for CQ is to implement the specification on top of actual quantum hardware. This is an ongoing effort, with some groups adopting and extending CQ for their research needs and hardware \cite{wennersteen2026}. Concurrently, we are in the process of implementing CQ on top of quantum resource management interface (QRMI) \cite{bacher2025, badts_examining_2026}. QRMI allows to incorporate quantum resources into existing workload managers (e.g. Slurm), discover those resources, acquire them and submit payload for execution. Given the growing number of supported quantum backends, this will enable us to offload quantum kernels and test CQ specification in its target environment, that is heterogeneous HPC systems. It should be noted that existing solutions do not fully respect closely-coupled hardware model assumed by CQ yet. Additional work with low-level quantum system controls (e.g. field-programmable gate arrays) is required. We envision that an access to quantum hardware across different vendors and platforms will shape further versions of CQ and available operations.

\section{Conclusion}

We have presented \emph{CQ}, a specification for a C-like API for quantum accelerated HPC, and \emph{CQ-SimBE} a reference implementation in C99 which offloads quantum computations to the statevector simulator, QuEST. Additionally, we have supplied the base interface with analogue extension enabling users to target platforms different than conventional gate-based quantum devices. We have further introduced native Fortran interface, lowering barriers to quantum accelerate existing scientific codes (written in Fortran). This combination of features places CQ in a unique spot in the space of programming interfaces.

We have demonstrated how this interface can be used to offload an example quantum Fourier transform kernel synchronously for multiple shots in a single call, and have provided the handles that allow the same kernel to be launched asynchronously, enabling both host and device to do useful work simultaneously. This ability to overlap host and accelerator compute is a key optimisation used in heterogeneous offloading and one that will need to be taken advantage of to enable efficient use of both quantum and classical resources. Indeed, recent work on quantum-enhanced Markov chain Monte Carlo \cite{layden_quantum-enhanced_2023} and quantum-enhanced parallel tempering \cite{ferguson_methods_2026, marshall_quantum-enhanced_2026}, highlights the possible benefits of the presented solution.

This is a first step on the road to enabling quantum offloading for HPC codes and provide the types of offloading interface the HPC community have become used to in GPU acceleration space. This work will enable HPC applications developed in both C and Fortran to more easily experiment with quantum algorithms, initially in quantum simulation but in future on dedicated quantum hardware.

\section*{Acknowledgments}

The authors acknowledge support from the Hub for Quantum Computing via Integrated and Interconnected Implementations (QCi3), funded by the EPSRC UK Quantum Technologies Programme under Grant Number EP/Z53318X/1. OTB and JR acknowledge funding from the Quantum Software Lab QATCH programme (UKRI4257).

\bibliographystyle{quantum} % Or your preferred style (e.g., apalike, IEEEtran)
\bibliography{CQ-Quantum}

@misc{lee23,
      title={Quantum Task Offloading with the {OpenMP API}},
      author={Joseph K. L. Lee and others},
      year={2023},
      eprint={2311.03210},
      archivePrefix={arXiv},
      primaryClass={cs.DC},
      url={https://arxiv.org/abs/2311.03210}, 
}

@repository{QuEST,
    title={{QuEST} v4},
    author={Tyson Jones and others},
    year={2025},
    code={https://github.com/QuEST-Kit/QuEST},
    version={4}
}

@article{jones19,
    title={{QuEST} and High Performance Simulation of Quantum Computers},
    author={Tyson Jones and Anna Brown and Ian Bush and Simon C. Benjamin},
    year={2019},
    journal={Scientific Reports},
    volume={9},
    pages={\,},
    number={10736},
    doi={10.1038/s41598-019-47174-9},
    url={https://doi.org/10.1038/s41598-019-47174-9}
}

@repository{CQ-Spec,
    title={{CQ}},
    author={Oliver Thomson Brown},
    year={2024},
    code={https://github.com/EPCCed/cq-spec}
}

@repository{CQ-SimBE,
    title={{CQ Simulated Backend}},
    author={Oliver Thomson Brown and Mateusz Meller and James Richings},
    year={2025},
    code={https://github.com/EPCCed/cq-simbe}
}

@article{Mohseni24,
      title={How to Build a Quantum Supercomputer: Scaling Challenges and Opportunities}, 
      author={Masoud Mohseni and others},
      year={2024},
      eprint={2411.10406},
      journal={arXiv},
      archivePrefix={arXiv},
      primaryClass={quant-ph},
      url={https://arxiv.org/abs/2411.10406}, 
}

@INPROCEEDINGS{BMH17,
  author={Britt, Keith A. and Mohiyaddin, Fahd A. and Humble, Travis S.},
  booktitle={2017 IEEE International Conference on Rebooting Computing (ICRC)}, 
  title={Quantum Accelerators for High-Performance Computing Systems}, 
  year={2017},
  publisher={IEEE},
  address={Washington, DC, USA},
  volume={},
  number={},
  pages={1-7},
  doi={10.1109/ICRC.2017.8123664}
}

@article{A-YCRK24,
doi = {10.1088/1361-6633/ad85f0},
year = {2024},
month = {10},
publisher = {IOP Publishing},
volume = {87},
number = {11},
pages = {116001},
author = {Au-Yeung, R and Camino, B and Rathore, O and Kendon, V},
title = {Quantum algorithms for scientific computing},
journal = {Reports on Progress in Physics}
}

@INPROCEEDINGS {SSRW23,
author = {Schulz, Martin and Schulz, Laura and Ruefenacht, Martin and Wille, Robert },
booktitle = { 2023 IEEE International Conference on Quantum Computing and Engineering (QCE) },
title = {Towards the Munich Quantum Software Stack: Enabling Efficient Access and Tool Support for Quantum Computers},
year = {2023},
volume = {},
ISSN = {},
pages = {399-400},
doi = {10.1109/QCE57702.2023.10301},
publisher = {IEEE Computer Society},
address = {Los Alamitos, CA, USA},
month = 9
}

@article{xacc_2020,
	doi = {10.1088/2058-9565/ab6bf6},
	year = 2020,
	month = {feb},
	publisher = {{IOP} Publishing},
	volume = {5},
	number = {2},
	author = {Alexander J McCaskey and others},
	pages = {024002},
	title = {{XACC}: a system-level software infrastructure for heterogeneous quantum{\textendash}classical computing},
	journal = {Quantum Science and Technology}
}

@INPROCEEDINGS{cuda-q,
  booktitle={2023 60th ACM/IEEE Design Automation Conference (DAC)}, 
  author={Kim, Jin-Sung and others},
  title={{CUDA} Quantum: The Platform for Integrated Quantum-Classical Computing}, 
  year={2023},
  address={San Francisco, CA, USA},
  publisher={ACM/IEEE},
  volume={},
  number={},
  pages={1-4},
  doi={10.1109/DAC56929.2023.10247886}
}

@INPROCEEDINGS{Kaya24,
  author={Kaya, Erc\"{u}ment and others},
  booktitle={2024 IEEE International Conference on Quantum Computing and Engineering (QCE)}, 
  title={{QPI}: A Programming Interface for Quantum Computers},
  publisher={IEEE},
  address={Montr\'{e}al, QC, Canada},
  year={2024},
  volume={02},
  number={},
  pages={286-291},
  doi={10.1109/QCE60285.2024.10293}
}

@article{meller25,
  doi = {10.22331/q-2025-12-02-1927},
  url = {https://doi.org/10.22331/q-2025-12-02-1927},
  title = {Programming tools for {A}nalogue {Q}uantum {C}omputing in the {H}igh-{P}erformance {C}omputing {C}ontext – {A} {R}eview},
  author = {Meller, Mateusz and Szeremi, Vendel and Brown, Oliver Thomson},
  journal = {{Quantum}},
  issn = {2521-327X},
  publisher = {{Verein zur F{\"{o}}rderung des Open Access Publizierens in den Quantenwissenschaften}},
  volume = {9},
  pages = {1927},
  month = dec,
  year = {2025}
}

@article{qiskit24,
      title={Quantum computing with {Q}iskit},
      author={Javadi-Abhari, Ali and others},
      year={2024},
      doi={10.48550/arXiv.2405.08810},
      eprint={2405.08810},
      journal={arXiv},
      archivePrefix={arXiv},
      primaryClass={quant-ph}
}

@article{pennylane22,
      title={PennyLane: Automatic differentiation of hybrid quantum-classical computations}, 
      author={Ville Bergholm and others},
      year={2022},
      eprint={1811.04968},
      archivePrefix={arXiv},
      journal={arXiv},
      primaryClass={quant-ph},
      url={https://arxiv.org/abs/1811.04968}, 
}

@article{Sivarajah21,
doi = {10.1088/2058-9565/ab8e92},
url = {https://dx.doi.org/10.1088/2058-9565/ab8e92},
year = {2020},
month = {nov},
publisher = {IOP Publishing},
volume = {6},
number = {1},
pages = {014003},
author = {Sivarajah, Seyon and others},
title = {$\mathrm{t}|\mathrm{ket}\rangle$: a retargetable compiler for {NISQ} devices},
journal = {Quantum Science and Technology}
}

@article{shehata26,
title = {Bridging paradigms: Designing for {HPC}-Quantum convergence},
journal = {Future Generation Computer Systems},
volume = {174},
pages = {107980},
year = {2026},
issn = {0167-739X},
doi = {https://doi.org/10.1016/j.future.2025.107980},
url = {https://www.sciencedirect.com/science/article/pii/S0167739X25002754},
author = {Amir Shehata and others}
}

@INPROCEEDINGS{sivarajah22,
  author={Sivarajah, Seyon and Heidemann, Lukas and Lawrence, Alan and Duncan, Ross},
  booktitle={2022 IEEE/ACM Third International Workshop on Quantum Computing Software (QCS)}, 
  title={Tierkreis: a Dataflow Framework for Hybrid Quantum-Classical Computing}, 
  year={2022},
  pages={12-21},
  publisher={IEEE/ACM},
  address={Dallas, TX, USA},
  doi={10.1109/QCS56647.2022.00007}
}

@repository{cunningham23,
  author       = {Will Cunningham and others},
  title        = {{AgnostiqHQ/covalent: v0.228.0-rc.0}},
  month        = sep,
  year         = 2023,
  publisher    = {Zenodo},
  version      = {v0.228.0-rc.0},
  doi          = {10.5281/zenodo.8369670},
  url          = {https://doi.org/10.5281/zenodo.8369670},
}

@misc{esposito25,
      title={SLURM Heterogeneous Jobs for Hybrid Classical-Quantum Workflows}, 
      author={Aniello Esposito and Utz-Uwe Haus},
      year={2025},
      eprint={2506.03846},
      archivePrefix={arXiv},
      primaryClass={cs.DC},
      url={https://arxiv.org/abs/2506.03846}, 
}

@inproceedings{qdmi,
    title = {{QDMI -- Quantum Device Management Interface: A Standardized Interface for Quantum Computing Platforms}},
    shorttitle = {{QDMI -- Quantum Device Management Interface}},
    booktitle = {IEEE International Conference on Quantum Computing and Engineering (QCE)},
    publisher={IEEE},
    author={Wille, Robert and others},
    address={Montr\'{e}al, QC, Canada},
    year = {2024},
    volume={02},
    number={},
    pages={573-574},
    doi={10.1109/QCE60285.2024.10411}
}

@website{qasm-gates,
title={{OpenQASM} Standard Library},
author={{OpenQASM Contributors}},
year={2025},
url={https://openqasm.com/language/standard\_library.html\#standard-library}
}

@repository{unity,
title={Unity Test},
author={Mike Karlesky and Mark {VanderVoord} and Greg Williams},
year={2025},
code={https://github.com/ThrowTheSwitch/Unity}
}

@article{parra-rodriguez20,
  title = {Digital-analog quantum computation},
  author = {Parra-Rodriguez, Adrian and others}, 
  journal = {Phys. Rev. A},
  volume = {101},
  issue = {2},
  pages = {022305},
  numpages = {12},
  year = {2020},
  month = {Feb},
  publisher = {American Physical Society},
  doi = {10.1103/PhysRevA.101.022305},
  url = {https://link.aps.org/doi/10.1103/PhysRevA.101.022305}
}

@article{layden_quantum-enhanced_2023,
	title = {Quantum-enhanced Markov chain Monte Carlo},
    author = {Layden, David and others}, 	
    volume = {619},
	rights = {2023 The Author(s), under exclusive licence to Springer Nature Limited},
	issn = {1476-4687},
	url = {https://www.nature.com/articles/s41586-023-06095-4},
	doi = {10.1038/s41586-023-06095-4},
	pages = {282--287},
	number = {7969},
    journal = {Nature},
	journaltitle = {Nature},
	publisher = {Nature Publishing Group},
	urldate = {2026-07-14},
	date = {2023-07},
    year = {2023},
	langid = {english},
}

@article{marshall_quantum-enhanced_2026,
	title = {Quantum-enhanced Markov Chain Monte Carlo for Combinatorial Optimization},
	url = {http://arxiv.org/abs/2602.06171},
	doi = {10.48550/arXiv.2602.06171},
	number = {{arXiv}:2602.06171},
	publisher = {{arXiv}},
    journal={arXiv},
	author = {Marshall, Kate V. and others},
	urldate = {2026-07-14},
	date = {2026-03-27},
	eprinttype = {arxiv},
	eprint = {2602.06171 [quant-ph]},
}

@article{ferguson_methods_2026,
	title = {Methods for non-variational heuristic quantum optimisation},
	url = {http://arxiv.org/abs/2602.01353},
	doi = {10.48550/arXiv.2602.01353},
	number = {{arXiv}:2602.01353},
	publisher = {{arXiv}},
    journal={arXiv},
	author = {Ferguson, Stuart and Wallden, Petros},
	urldate = {2026-07-14},
	date = {2026-02-01},
	eprinttype = {arxiv},
	eprint = {2602.01353 [quant-ph]},
}

@article{badts_examining_2026,
	title = {Examining {QRMI} as a Unified Interface for {Quantum-HPC} Integration},
	url = {https://arxiv.org/abs/2607.19591},
	doi = {10.48550/arXiv.2607.19591},
	number = {{arXiv}:2607.19591},
	publisher = {{arXiv}},
    journal={arXiv},
	author = {Badts, Thomas and others},
	urldate = {2026-08-18},
	date = {2026-08-08},
	eprinttype = {arxiv},
	eprint = {2607.19591},
}

@article{vazquez-perez24,
	title = {{QPU} integration in {OpenCL} for heterogeneous programming},
	volume = {80},
	issn = {1573-0484},
	url = {https://doi.org/10.1007/s11227-023-05879-9},
	doi = {10.1007/s11227-023-05879-9},
	pages = {11682--11703},
	number = {8},
	journaltitle = {J Supercomput},
	author = {Vázquez-Pérez, Jorge and Piñeiro, César and Pichel, Juan C. and Pena, Tomás F. and Gómez, Andrés},
	urldate = {2026-09-24},
	date = {2024-05-01},
	langid = {english},
}

@article{gazda23,
	title = {A pragma based {C++} framework for hybrid quantum/ classical computation},
	rights = {Creative Commons Attribution 4.0 International},
	url = {https://arxiv.org/abs/2309.02605},
	doi = {10.48550/ARXIV.2309.02605},
	publisher = {{arXiv}},
    journal={arXiv},
	author = {Gazda, Arnaud and Koska, Oceane},
	urldate = {2026-09-24},
	date = {2023},
	note = {Version Number: 3},
}

@misc{dobler2025,
      title={Extending the OmpSs-2 Programming Model for Hybrid Quantum-Classical Programming}, 
      author={Philip Döbler and David Álvarez and Lucas J. Menger and Thomas Lippert and Vicenç Beltran and Manpreet Singh Jattana},
      year={2025},
      eprint={2502.21104},
      archivePrefix={arXiv},
      primaryClass={cs.ET},
      url={https://arxiv.org/abs/2502.21104}, 
}

@inproceedings{swastik2025,
author = {Mittal, Swastik and Mahesh, Atulya and Mueller, Frank},
title = {{OpenMP-Q}: Quantum Task Offloading in {OpenMP}},
year = {2025},
isbn = {978-3-032-06342-7},
publisher = {Springer-Verlag},
address = {Berlin, Heidelberg},
url = {https://doi.org/10.1007/978-3-032-06343-4_6},
doi = {10.1007/978-3-032-06343-4_6},
booktitle = {OpenMP: Balancing Productivity and Performance Portability: 21st International Workshop on OpenMP, IWOMP 2025, Charlotte, NC, USA, October 1–3, 2025 Proceedings},
pages = {81–95},
numpages = {15}
}

@repository{wennersteen2026,
title={CQ-SimBE: Payload Export Fork},
author={Aleksander Wennersteen},
year={2026},
code={https://github.com/awennersteen/cq-simbe/tree/aw/cq-exports}
}

@article{gokhale2026,
    author  = {Gokhale, Shreeya Sanjeev and Dhote, Rishit Ravi and Delhibabu, Radhakrishnan},
    title = { A review of quantum machine learning algorithms, applications, and emerging advantages},
    journal = {Discover Computing},
    volume = {29},
    number = {226},
    url = {https://doi.org/10.1007/s10791-026-10085-1},
    doi = {10.1007/s10791-026-10085-1},
    year = {2026}
}

@book{nelson1981,
  title={Remote procedure call},
  author={Nelson, Bruce Jay},
  year={1981},
  publisher={Carnegie Mellon University},
  url={https://dl.acm.org/doi/abs/10.5555/910306}
}

@article{birrell1984,
author = {Birrell, Andrew D. and Nelson, Bruce Jay},
title = {Implementing Remote procedure calls},
year = {1983},
issue_date = {October 1983},
publisher = {Association for Computing Machinery},
address = {New York, NY, USA},
volume = {17},
number = {5},
issn = {0163-5980},
url = {https://doi.org/10.1145/773379.806609},
doi = {10.1145/773379.806609},
journal = {SIGOPS Oper. Syst. Rev.},
month = oct,
pages = {3},
numpages = {1}
}

@misc{bacher2025,
      title={Quantum resources in resource management systems}, 
      author={Utz Bacher and Mark Birmingham and Christopher D. Carothers and Andrew Damin and Carlos D. Gonzalez Calaza and Ashwin Kumar Karnad and Stefano Mensa and Matthieu Moreau and Aurelien Nober and Munetaka Ohtani and Max Rossmannek and Philippa Rubin and M. Emre Sahin and Oscar Wallis and Amir Shehata and Iskandar Sitdikov and Aleksander Wennersteen},
      year={2025},
      eprint={2506.10052},
      archivePrefix={arXiv},
      primaryClass={quant-ph},
      url={https://arxiv.org/abs/2506.10052}, 
}

\newpage

\appendix

\section{Details of the CQ Execution Handle Interface}
\label{appendix:exec_interface}

The interface for interacting with CQ execution handle is outlined below in \cref{lst:cq-exec}. The interface consists of resource management functions for allocation and deallocation of the handle as well as a minimal set of getter functions, required for meaningful usage of CQ with non-blocking offloading pattern.

\begin{lstlisting}[label={lst:cq-exec}, language=C, caption={CQ interface for interacting with execution handle. These include: allocation and deallocation function which return status code, as well as, getters for important information, such as number of completed or expected shot, required for asynchronous offloading.}, captionpos=b]
int alloc_exec(struct exec ** ehp, cstate** cr, 
    const size_t NMEASURE, const size_t NSHOTS);
int free_exec(struct exec ** ehp);

size_t exec_id(struct exec * const eh);
bool exec_is_init(struct exec * const eh);
bool exec_completed(struct exec * const eh);
bool exec_halted(struct exec * const eh);
int exec_status(struct exec * const eh);
size_t exec_completed_shots(struct exec * const eh);
size_t exec_expected_shots(struct exec * const eh);
\end{lstlisting}

\section{Details of the CQ Fortran Interface}
\label{appendix:fort_interface}

For completeness, in \cref{cq-fort-abstr-interface}, we present Fortran implementation of the abstract interface for functions specifying CQ kernels.
\begin{lstlisting}[label={cq-fort-abstr-interface}, language=Fortran, caption={Fortran-defined, abstract interface of a quantum kernel. This is a prototype for the quantum kernels.}, captionpos=b]
abstract interface
  function qkern(NQUBITS, qr, NMEASURE, cr, reg) result(status) bind(C)
    use iso_c_binding, only: c_int, c_size_t, c_short
    import :: qubit, qkern_map
    implicit none
    
    integer(c_size_t), value ::NQUBITS
    type(qubit), value :: qr
    integer(c_size_t), value :: NMEASURE
    integer(c_short), intent(inout) :: cr(0:NMEASURE)
    type(qkern_map), value :: reg
    integer(c_int) :: status
  end function qkern
end interface
\end{lstlisting}

Using this interface, a concrete implementation of a kernel is shown below in \cref{lst:cq-fort-qft}.

\begin{lstlisting}[label={lst:cq-fort-qft}, language=Fortran, caption={Example implementation of a QFT using CQ-based Fortran interface. The quantum kernel definition follows standard CQ-outlined structure, where, first, kernel has to be registered via preprocessor macro, followed by quantum device operations, calls to sub-kernels and optionally a measurement.}, captionpos=b]
function plus_state_qft(NQUBITS, qr, NMEASURE, cr, reg) bind(C)
  implicit none
  integer(kind=8), value :: NQUBITS
  type(qubit), value :: qr
  integer(kind=8), value :: NMEASURE
  integer(kind=2), intent(inout) :: cr(0:NMEASURE)
  type(qkern_map), value :: reg
  integer(kind=8) :: i
  integer(kind=8) :: STATE_IDX = 0
  integer :: status
  CQ_REGISTER_KERNEL("plus_state_qft", reg)
  status = cq_set_qureg(qr, STATE_IDX, NQUBITS)

  do i = 0, NQUBITS-1
    status = cq_hadamard(qr, i)
  end do

  call qft(NQUBITS, qr)
  status = cq_measure_qureg(qr, NQUBITS, cr)
end function plus_state_qft
\end{lstlisting}

\end{document}